\documentstyle[referee,graphicx]{mn}

\newif\ifAMStwofonts

\newfont{\Varepsilon}{cmsy10 at 18.0pt}
\newcommand{\msun}{M_{\odot}}

\ifoldfss
  \ifCUPmtlplainloaded \else
    \NewTextAlphabet{textbfit} {cmbxti10} {}
    \NewTextAlphabet{textbfss} {cmssbx10} {}
    \NewMathAlphabet{mathbfit} {cmbxti10} {} 
    \NewMathAlphabet{mathbfss} {cmssbx10} {} 
  \fi
  \ifAMStwofonts
    \ifCUPmtlplainloaded \else
      \NewSymbolFont{upmath} {eurm10}
      \NewSymbolFont{AMSa} {msam10}
      \NewMathSymbol{\upi}     {0}{upmath}{19}
      \NewMathSymbol{\umu}     {0}{upmath}{16}
      \NewMathSymbol{\upartial}{0}{upmath}{40}
      \NewMathSymbol{\leqslant}{3}{AMSa}{36}
      \NewMathSymbol{\geqslant}{3}{AMSa}{3E}

    \fi
  \fi
\fi 

\ifnfssone
  \newmathalphabet{\mathit}
  \addtoversion{normal}{\mathit}{cmr}{m}{it}
  \addtoversion{bold}{\mathit}{cmr}{bx}{it}
  \newmathalphabet{\mathbfit} 
  \addtoversion{normal}{\mathbfit}{cmr}{bx}{it}
  \addtoversion{bold}{\mathbfit}{cmr}{bx}{it}
  \newmathalphabet{\mathbfss} 
  \addtoversion{normal}{\mathbfss}{cmss}{bx}{n}
  \addtoversion{bold}{\mathbfss}{cmss}{bx}{n}
  \ifAMStwofonts
    \ifCUPmtlplainloaded \else
      \UseAMStwoboldmath
      \makeatletter
      \new@mathgroup\upmath@group
      \define@mathgroup\mv@normal\upmath@group{eur}{m}{n}
      \define@mathgroup\mv@bold\upmath@group{eur}{b}{n}
      \edef\UPM{\hexnumber\upmath@group}
      \new@mathgroup\amsa@group
      \define@mathgroup\mv@normal\amsa@group{msa}{m}{n}
      \define@mathgroup\mv@bold\amsa@group{msa}{m}{n}
      \edef\AMSa{\hexnumber\amsa@group}
      \makeatother
      \mathchardef\upi="0\UPM19
      \mathchardef\umu="0\UPM16
      \mathchardef\upartial="0\UPM40
      \mathchardef\leqslant="3\AMSa36
      \mathchardef\geqslant="3\AMSa3E
    \fi
  \fi
\fi 

\ifnfsstwo
  \DeclareMathAlphabet{\mathbfit}{OT1}{cmr}{bx}{it}
  \SetMathAlphabet\mathbfit{bold}{OT1}{cmr}{bx}{it}
  \DeclareMathAlphabet{\mathbfss}{OT1}{cmss}{bx}{n}
  \SetMathAlphabet\mathbfss{bold}{OT1}{cmss}{bx}{n}
  \ifAMStwofonts
    \ifCUPmtlplainloaded \else
      \DeclareSymbolFont{UPM}{U}{eur}{m}{n}
      \SetSymbolFont{UPM}{bold}{U}{eur}{b}{n}
      \DeclareSymbolFont{AMSa}{U}{msa}{m}{n}
      \DeclareMathSymbol{\upi}{0}{UPM}{"19}
      \DeclareMathSymbol{\umu}{0}{UPM}{"16}
      \DeclareMathSymbol{\upartial}{0}{UPM}{"40}
      \DeclareMathSymbol{\leqslant}{3}{AMSa}{"36}
      \DeclareMathSymbol{\geqslant}{3}{AMSa}{"3E}
    \fi
  \fi
\fi 

\ifCUPmtlplainloaded \else
  \ifAMStwofonts \else 
    \def\upi{\pi}
    \def\umu{\mu}
    \def\upartial{\partial}
  \fi
\fi

\title[Stellar abundance anomalies and H shell instabilities]
{An attempt to model globular cluster red giant abundance anomalies 
with a simulated hydrogen shell instability}
\author[B. B. Messenger and J. C. Lattanzio]
       {B. B. Messenger and J. C. Lattanzio\\
        Department of Mathematics and Statistics, PO Box 28M, 
        Monash University, 3800, Australia}
\date{Accepted ???.
      Received ???;
      in original form 2000 June ??}

\pagerange{\pageref{firstpage}--\pageref{lastpage}}
\pubyear{1994}

\begin{document}

\maketitle

\label{firstpage}

\begin{abstract}
It has been suggested that the anomalous Na, Mg and Al observed in Globular Cluster
Red Giant stars could be the result of a thermally unstable hydrogen shell.
Currently accepted reaction rates indicate that temperatures of approximately 70-75 million K are required to produce the observed enhancements in Na and Al along
with depletions in Mg.

The work presented here attempts to model the H shell instability by a simple
mechanism of altering the energy production in the region of the H shell. We
show that even extreme cases only give rise to small intermittent temperature
increases that have minimal affect on the surface abundances. Full evolutionary
modelling incorporating this technique simply accelerates the evolution of the
RGB phase producing the same surface abundances as other models but at an
earlier time.  We conclude that, unless hydrogen shell instabilities manifest 
themselves quite differently, they are unlikely to lead to the required 
temperatures and alternative explanations of the abundance anomalies are more 
promising.
\end{abstract}


\section{Introduction}
As long ago as 1947, Popper \shortcite{Popper:47a} had noted that the 
globular cluster
red giant L199 in M13 was CN-strong. Further work by Harding
\shortcite{Harding:62a}, Osborn \shortcite{Osborn:71a} and Hesser, Hartwick and McClure
\shortcite{Hesser:76a} confirmed that this anomaly is common. So for over fifty
years it has been known that the surface abundances of elements such as C and N
can vary from one red giant star to 
another within an individual globular
cluster.  It has become clear in the intervening years, however, that these
anomalies are not restricted to C and N:~heavier elements such as O, Na, Mg
and Al also show star to star variations. Most other elements studied show
little variation, and the degree of variation of C, N, O etc. differs from
cluster to cluster and is much less apparent in field stars \cite{Langer:92a}. 

Observations by many groups, including Kraft et al. \shortcite{Kraft:97a}, 
Smith and Kraft \shortcite{Smith:96a}, Langer \shortcite{Langer:92a}, Shetrone
\shortcite{Shetrone:96a} \shortcite{Shetrone:96b} \shortcite{Shetrone:97a}, 
have identified this abundance anomaly problem and the most commonly 
suggested 
explanation for such high Na and Al are 
the deep mixing \cite{Sweigart:79a} and primordial enhancement
\cite{Cottrell:81a} hypotheses. The deep mixing hypothesis suggests 
that the rotation of the star can give rise to meridional circulation 
currents that in turn create a mixing zone in the radiative region 
separating the top of the hydrogen burning shell (HBS) and the base 
of the convective envelope (BCE). This would allow elements processed 
in the HBS to be mixed to the surface as the star proceeds up the RGB. 
The primordial enhancement hypothesis suggests that the stars are born
 with these anomalies already present. The stars are formed from material 
ejected by intermediate mass ($\simeq 3-10 \msun$) AGB stars from an earlier 
epoch. Without going into the full mechanisms of each process, either
could in principle explain the observed anomalies, but modelling and observations
have 
indicated that probably both processes occur to some extent.

Langer, Hoffman and Zaidins \shortcite{Langer:97a} 
suggested that the Na and Al abundance
anomaly problem may be solved if the H shell reached higher temperatures for at
least part of its lifetime whilst on the RGB. The idea stems from work by Von
Rudloff and Vandenburg \shortcite{VonRudloff:88a} and even earlier work by Bolton and
Eggleton \shortcite{Bolton:73a}. Von Rudloff and Vandenburg \shortcite{VonRudloff:88a}
found that the H shell of a
non-rotating small mass red giant model was stable but only marginally. A
rotating model could be unstable. The link with rotation is particularly
encouraging because this mechanism could explain both the deep mixing and the
thermally unstable shell.
The work performed by Langer et al \shortcite{Langer:97a}
shows that a thermally unstable H shell
would allow for the greater production of $^{27}$Al at the expense of $^{24}$Mg
and that there would also be extra $^{23}$Na available for mixing to the
surface. 

The conclusion was that if the H shell reached temperatures of approximately
70-75 million K then the material mixed to the surface would 
produce a composition
in close
agreement with observation. In fact temperatures of this order and no lower
would be required to match the Na observations because at the lower
temperatures too much $^{23}$Na would be produced. Also it is only at these
temperatures that significant $^{27}$Al is produced and $^{24}$Mg is only
destroyed at temperatures above 70 million K. Therefore, to have any surface
depletions in $^{24}$Mg the temperature of the H shell has
to reach at least 70 million K unless primordial changes are invoked or the
accepted reaction rates are altered. The preferred temperature range was 72-73
million K (see section 2 of Langer et. al \shortcite{Langer:97a}). This is the
temperature at which they believed all the observed surface abundance anomalies
could be matched.

As the structure of a model is heavily
dependent upon the temperature profile any
change to the temperature must be included in the evolution modelling.
The following sections discuss models where temperatures in the H shell
are artificially increased by decreasing the energy generation rate. 
The next section
looks at oscillating temperature rises in the HBS; i.e. a pulsating model where
each pulse immediately follows the previous pulse. In the following section 
pulses
in the temperature of the HBS are created with a time gap between each pulse. 

\section{Computer Codes}
\label{Codes Used}
The calculations reported here make use of two separate
computer codes:~the Monash/Mt~Stromlo stellar 
evolution
code, with which we model the evolution of stars from contraction to the main 
sequence up to the RGB tip, and a post-processing nucleosynthesis code 
\cite{Cannon:90a}, with which we model the abundances within the star for the 
same epoch. The nucleosynthesis code uses the independent variables from the 
evolution code for its structural basis, with typically every sixth evolution 
model being used. The full details of how these two codes work and interface 
can be found in Messenger \shortcite{Messenger:00a}. 

The postulated 
deep mixing has been included in the nucleosynthesis code,
and is described by three parameters - the depth, the speed and 
the time of onset of the deep mixing. 
We measure the depth in the same way as
Wasserburg, Boothroyd and Sackmann 
\shortcite{Boothroyd:97a}, by mixing down to 
a temperature $\triangle \log T$ 
above the base of the HBS, which is defined as the first point (moving outward from
the center) to have a non-zero hydrogen abundance. This mixing commences once
the HBS burns through the discontinuity left from the first dredge-up episode earlier 
in the evolution. Prior to this time it is assumed that this molecular weight 
discontinuity will inhibit mixing: once it is removed, the mixing is assumed to begin.
Finally, we found (in agreement with previous researchers) that provided the speed
of mixing was above a small critical value, it made almost no difference to the
abundance patterns. We used a mixing speed of $10^{-4}\msun$/year in all
calculations reported here.

In the work of Messenger 
\shortcite{Messenger:00a} it was found that varying the mixing depth as the 
star ascends the RGB gives better agreement with observations of Na. 
Therefore this type of modelling is adopted 
in this paper,
and the mixing depth was 
changed linearly from an initial setting $\triangle \log T=0.18$ to 
$\triangle \log T=0.03$  as the star ascended the RGB.

\section{Increased H shell Temperatures}
\label{increased_H_shell:sec}
\subsection{Algorithm}

To increase the temperature of the H shell it is not simply a matter of arbitrarily adding temperature to the
mesh points within the shell. The temperature is one of the dependent
structure variables and affects other
variables such as the density and pressure through the equation of state. To self-consistently increase the temperature we chose to change the energy generation of the shell by increasing or decreasing the energy production rate for each point if it is within the boundaries of the H
shell.

An increase of the energy generation rate causes the star to expand and 
the temperature to decrease. A reduction in the energy generation rate causes
the opposite; the star contracts and the temperature increases.
Therefore it was necessary to decrease the energy rate from H burning in
order to artificially create a temperature increase.

A factor for the energy generation rate reduction,
designated as $g$, was introduced.  This reduction is only
initiated after the H shell passes through the $\mu$-discontinuity remaining 
from the first dredge-up episode. The desired nuclear energy 
generation rate at a point in
the H shell (when this feature is switched on) is

\begin{equation}
\indent \indent \indent \indent \epsilon_{used} = \epsilon_{rate}(1-g)
\end{equation}

However, it is not simply a matter of reducing the energy generation rate 
over the full shell. This would
create a large jump in energy generaqtion
from the point just below the shell and also
 from the point at the top of the
shell.\footnote{We define the top of the shell to be the point that first 
has a hydrogen abundance 0.98 of the surface
abundance moving from the centre of the model outwards.} It is necessary 
to linearly interpolate from 0 to the
value of $g$ over a number of mesh points at the bottom of the shell and 
from $g$ to 0 at the top of the shell.

\begin{figure}
\includegraphics[width=0.6\textwidth,viewport=150 80 660 470]
{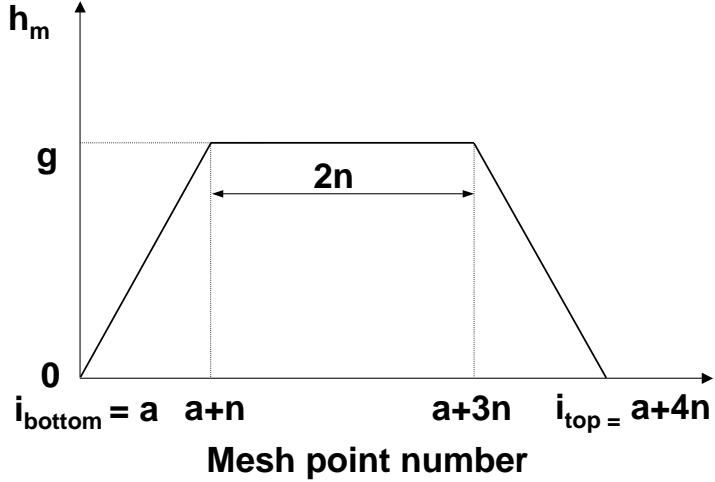}
\caption{\label{trap_energy_shell_mass:fig} Basic energy generation 
rate reduction algorithm. The mesh-based
energy generation reduction factor $h_{m}$ is linearly increased from zero 
to $g$ over $n$ mesh
points, held at $g$ for $2n$ mesh points and then linearly 
decreased to zero
over a further $n$ mesh points.}
\end{figure}

Let $h_{m}$ be the mesh-based linear interpolation factor.
This varies from zero at the bottom of the HBS to a maximum of $g$ over $n$
mesh-points, remains at this maximum over $2n$ mesh points, and
then decreases linearly over the next $n$ mesh points to a value of zero again.
This is shown in Figure~\ref{trap_energy_shell_mass:fig}.

Once $h_{m}$ has been determined the energy generation rate remaining at 
any point within the H shell (as long
as the H shell instability feature is switched on) is

\begin{equation}
\indent \indent \indent \indent \epsilon_{used} = \epsilon_{rate}(1-h_{m})
\end{equation}

This alone caused a temporal discontinuity in the models and subsequently
created convergence problems. Remember that this arbitrary temperature rise is
non-physical. 
To avoid these convergence
problems it was necessary to smoothly introduce 
the reduced energy generation rate in time
as well as in mass (i.e. to gradually turn on the energy reduction).
We introduce the reduction rate gradually over time with a 
 parameter $h_{t}$ which is linearly increased from 0 to 1 over 20 
timesteps. Thus the energy generation rate within the H shell is

\begin{equation}
\indent \indent \indent \indent \epsilon_{used} = \epsilon_{rate}(1-h_{m}h_{t})
\end{equation}

Note that parametrisations in terms of mesh point (rather than mass) and model number
(rather than time) are purely numerical conveniences. In reality, of course, these
variations should be dependent on mass and time, but without any identified
mechanism for the instability, and in view of the exploratory nature of the
calculations reported here, we believe this simplification is justified and
in no way affects our conclusions, which depend only on the shell temperature.

\subsection{Results}
All calculations reported here are carried out for a star of $0.8\msun$ with
a metallicity of $Z=0.0005$. Parameters relevant to the deep mixing were given 
above.

\begin{figure}
\includegraphics[height=0.8\textwidth,angle=-90]{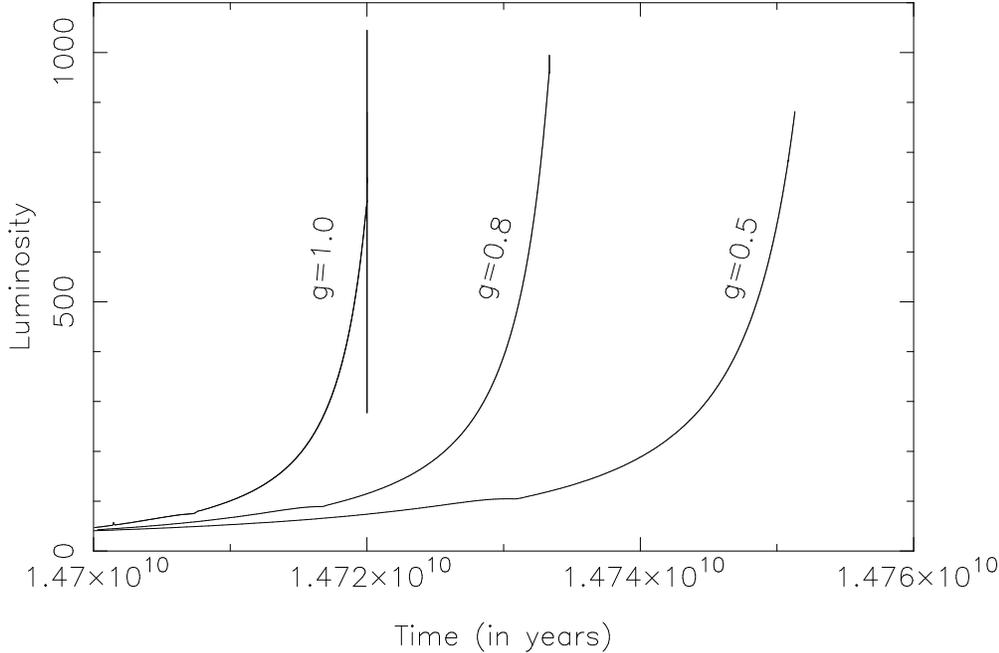}
\caption{\label{basic_e_removal:fig} Model for a constant energy generation
rate reduction. As can be
seen the greater the reduction of the nuclear burning energy rate, the more
rapid the evolution of the model. For $g=0.5$ the time spent on the RGB (since first dredge-up) is
approximately 7.3\% less than a star where $g=0$. Similarly for $g=0.8$ the time
spent is approximately 12\% less and for $g=1.0$ the time spent is approximately
16\% less.}
\end{figure}

The temperature in the H shell is determined
by the requirement that the energy production rate 
be high enough to support the outer layers of the star. With a decreased rate of 
energy generation at a given temperature, the star's response is to increase 
the temperature in the H shell. As the star evolves up the
RGB, the H shell progresses outwards and thins. The higher temperatures mimic
this process and the final result is an acceleration of the star's evolution.
This is shown in figure \ref{basic_e_removal:fig}, where plots of luminosity
against time for differing values of $g$ are shown. The greater the
energy rate reduction, the quicker the star evolves to the RGB tip.

This result will not assist in the abundance anomaly problem because
any enhancements/depletions will still occur to about the same degree 
as before but just at an earlier epoch. The higher temperatures within the 
H shell mimic normal evolution but reach it at an earlier stage. 
All this manages to do is
force the model to the He flash more quickly. No extra Na, Mg or Al would be
produced in this situation. in fact, with a reduced timescale for the 
burning, the effect may be the opposite of what is required, but 
in any event it will be quantitatively negligible.

\section{Intermittent Temperature Rises}

Langer et. al. \shortcite{Langer:97a} suggested that the H shell is a thermally
unstable region and that under certain conditions the shell may give rise to
higher temperatures than standard physics suggests. The modelling performed in
the previous section 
is not that of a thermally
unstable region,
but just  a consistently higher
temperature. To model a thermal instability the energy generation rate
modification must not be steady but should be periodic or at least variable. A
small modification to the algorithm of section \ref{increased_H_shell:sec} was
therefore made to simulate a thermally unstable H shell.

\begin{figure}
\includegraphics[height=0.8\textwidth, angle=-90]{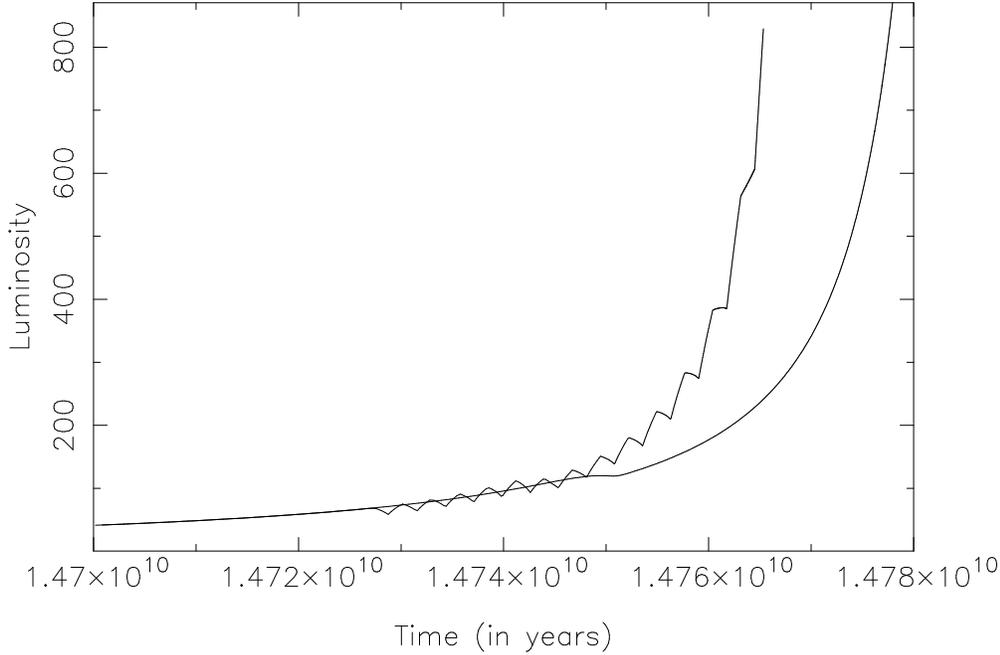}
\caption{\label{time_pulse_no_gaps_on_basic:fig} A periodic energy generation rate reduction model and a standard model with normal energy production.}
\end{figure}

The simplest way to do this is to continue to use the time based rate
reduction algorithm but just make it periodic. 
Here the energy generation rate reduction was introduced over five timesteps,
held at the peak for 20 timesteps and then decreased over another five
timesteps. This was then repeated until the RGB tip was reached. For an initial
test and to determine how pulsation may affect convergence there was no gap
between each pulse. Apart from the increase and decrease of each pulse the
affect would be identical to the continuous algorithm described in the
previous section. We would therefore expect little change from the model for
the previous algorithm and that is exactly the case. Figure
\ref{time_pulse_no_gaps_on_basic:fig} which has  $g=0.8$ 
shows that the result of the above algorithm is
to accelerate the evolution just as it did for the case of a continuous H shell
temperature rise. The problem with the above approach is that the decrease in
the energy generation rate causes the shell to contract and temperatures to
rise. This shifts the star to a different evolutionary track. Once the energy
generation rate returns to normal values the shell begins to expand again
but there is not enough time for the star
to settle back to its original (or close to original) evolutionary track
before the next pulse begins.
The end result is that the star shifts from its original
evolutionary track and diverges
from it henceforth.

If we do not want to significantly alter the evolution of the star, then
it is necessary to introduce a time delay
between pulses which is long enough for the star to
resettle to its original track.

We use  a simple method, but it is efficient and can test the concept of a H
shell instability without the introduction of large amounts of complex code
which is hard to justify since there is no confirmed mechanism for 
the instability.  The purpose of our algorithm is to test the 
principle of a periodic H shell instability, not the details.
We use a technique similar in principle to that shown in figure 1 
but we introduce periodicity and time
delays between pulses. We firstly select the desired number of pulses 
between FDU and the tip of the giant branch. We thus determine the
inter-pulse duration and distribute the pulses evenly over 
the giant branch life-time. After some experimentation we decided
to use 20 pulses because this allows the star to settle back to 
its standard evolutionary track and it should illustrate any differences 
likely to occur due to the pulses. We then take as a characteristic
time-scale the time-step used to construct the current model, which we
denote as $dt$.
Next, the duration of a pulse is taken
as $30dt$, comprising  $5 dt$ over which we linearly increase 
$h_t$ from zero to unity, then 
$20 dt$ using $h_t=1$, and a further
$5 dt$ when $h_t$ is linearly decreased back to zero.
Of course, the time-steps used by the code vary,
and are not all equal to $dt$ during the pulse, but $dt$ gives a convenient
timescale for the test. Also, because the time-steps vary with evolution
along the giant branch, the $dt$ used for each of the 20 pulses
will be slightly different, but this is not important for our exploratory
calculations.

\begin{figure}
\includegraphics[height=0.8\textwidth, angle=-90]{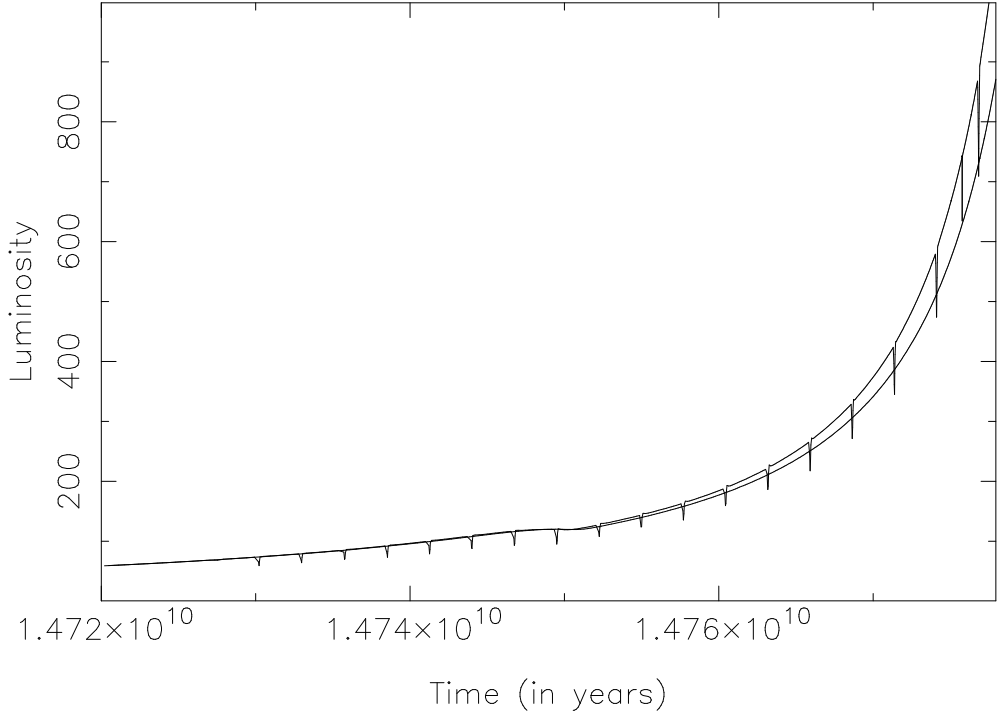}
\caption{\label{none_and_pulse_e_removal:fig}Periodic energy generation rate reduction model and a standard model with no energy generation rate
modifications. Here the evolution is not accelerated as much because there is enough time between pulses to allow the
model to resettle on to its original evolutionary track. The model does deviate as it approaches the RGB tip
but this could be avoided by selecting a smaller number of pulses.}
\end{figure}

The result of the introduction of this algorithm 
is shown in figure \ref{none_and_pulse_e_removal:fig}.
The points that should be
noted are the decrease in the luminosity shown during a pulse accompanied by a
temperature increase and the fact that the curve and hence the model settles
back to its original track. It is only at the latter stages of the RGB evolution that
the curves begin to diverge.

\begin{figure}
\includegraphics[height=0.8\textwidth, angle=-90]{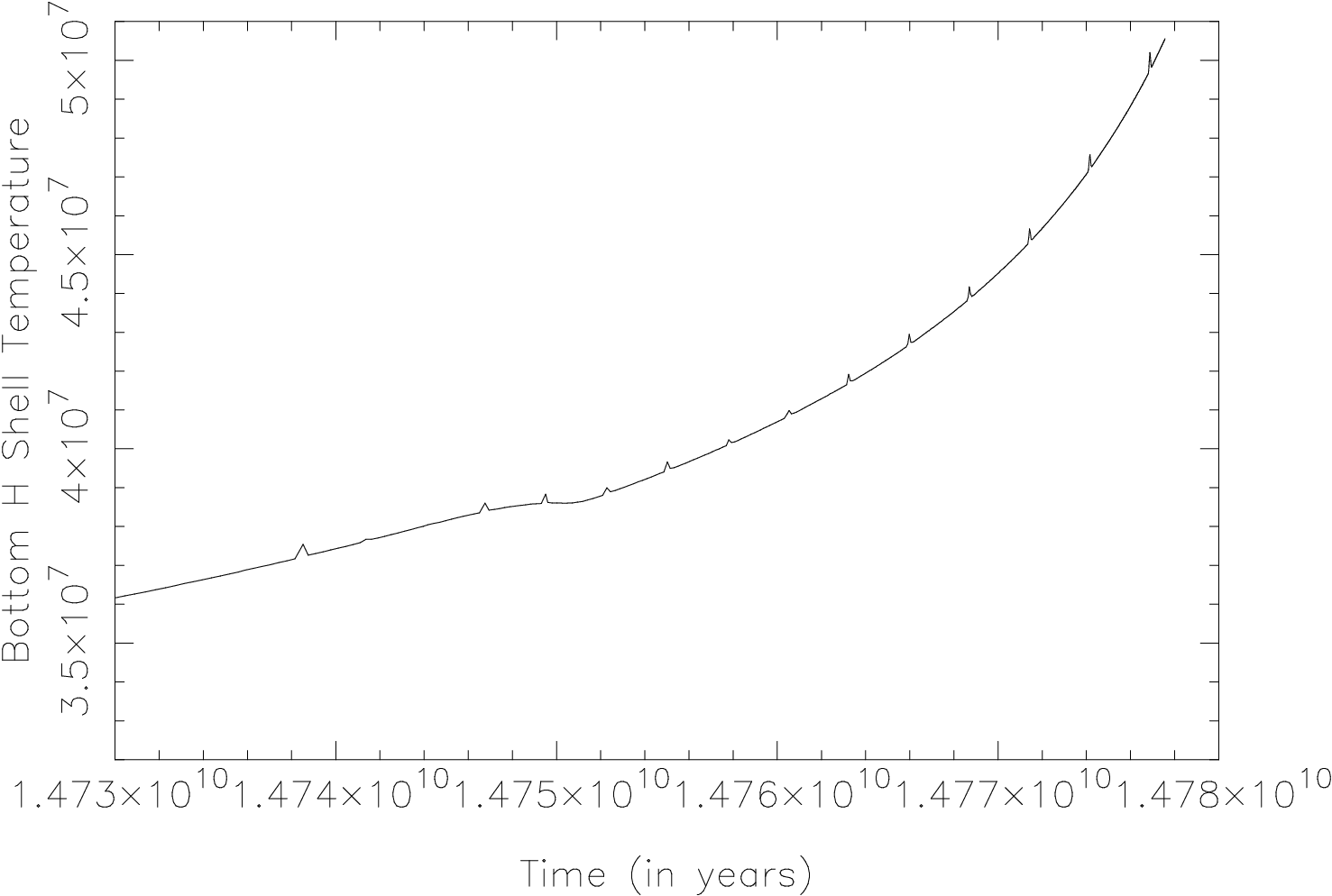}

\caption{\label{BHS_0.5:fig} The temperature at the base of the hydrogen shell for a periodic energy generation rate
reduction model with $g$=0.5.}

\end{figure}

The conclusion drawn from this model is that the H shell temperature can be
increased intermittently without significantly affecting the evolution of the
star. But how will it affect the nucleosynthesis of the model? Is the
temperature increase high enough and long enough to produce significant amounts
of $^{27}$Al and reduce $^{24}$Mg as originally hoped? It is necessary to look
at the temperatures within the H shell to see whether temperatures of
approximately 70 million K are reached, the temperature at which $^{27}$Al
enhancements are reasonable and  $^{24}$Mg depletions begin. Figure
\ref{BHS_0.5:fig} shows the temperature at the base of the H shell as the star
ascends the RGB for an energy generation rate reduction of 50\% ($g=0.5$). 
The peak
temperature reached by any of the pulses is only increased very slightly, well short
of the required temperatures for $^{24}$Mg destruction. Even a 100\% reduction
of the nuclear burning energy generation rate ($g=1.0$) 
during a pulse, as shown in
figure \ref{BHS_1.0:fig}, cannot produce high enough temperatures within the
shell for the processes discussed here. For a test case, some of the
gravitational and neutrino energy generation rates were also reduced
($g=1.5$). This
created a minor increase in the temperature (see figure \ref{BHS_1.5:fig}), but
the peak temperature still fell short of the required temperatures.

\begin{figure}
\includegraphics[height=0.8\textwidth, angle=-90]{BHS_1.0.ps}
\caption{\label{BHS_1.0:fig} The temperature at the base of the hydrogen shell1 for a periodic energy generation rate
reduction model with $g$=1.0.}
\end{figure}

\begin{figure}
\includegraphics[height=0.8\textwidth, angle=-90]{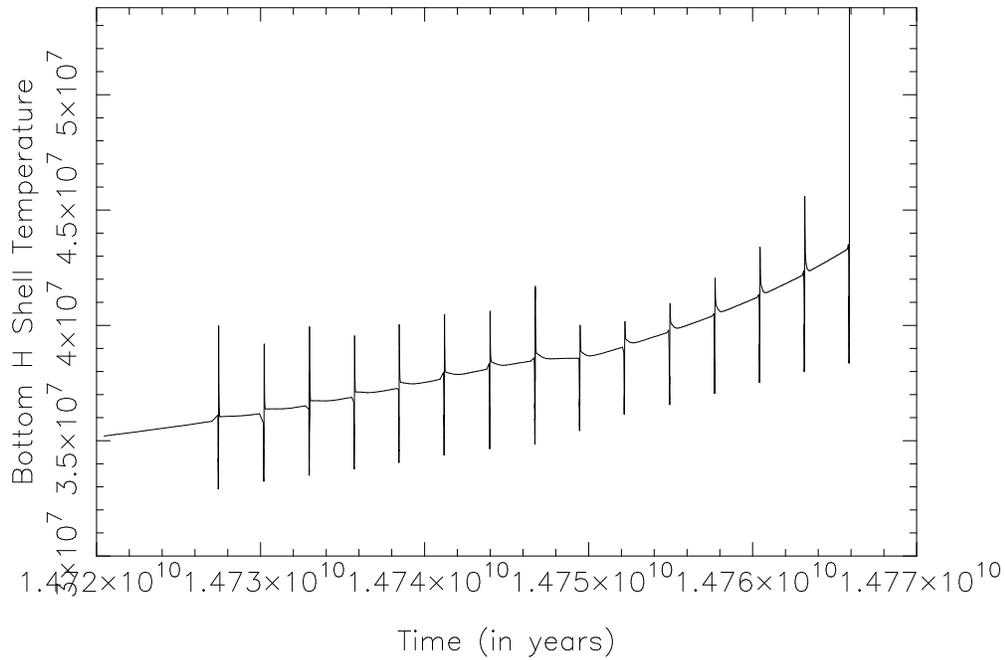}

\caption{\label{BHS_1.5:fig} The temperature at the base of the hydrogen shell for a periodic energy
generation rate reduction model with $g$=1.5.
(Thus removing all the burning energy, some, possibly all, the gravitational energy and the total energy may even be negative.)}

\end{figure}

Also if there are to be enhancements from this temperature increase, they must
reach such temperatures early enough for significant composition differences
to be seen in the envelope.
The peak temperatures from these models are only seen 
during the very latest stages
of RGB evolution, probably not early enough to match the 
data from M13 for example.

It seems that the required temperature increase cannot be produced via this
method. The suggested instability has some desirable features but to self
consistently produce the required temperatures proves difficult: our simulated
instability does not succeed, and the question of whether a genuine instability
would succeed remains open until such a mechanism can be found, and 
consistently modelled.
We are therefore reluctant to dismiss this
possibility, based on the technique attempted in this paper.
Also a genuine instability may be different to what we have modelled here and
may create higher temperatures in the HBS.

Despite this, temperature rises are seen whilst the pulse is 
active, and these could alter the
nucleosynthesis. It would not be expected that any $^{24}$Mg would be
destroyed at these temperatures, 
but $^{27}$Al production starts at lower temperatures and peak $^{23}$Na 
production at even lower temperatures. Even though the temperature pulses 
are not high enough to test the
hypothesis espoused by Langer et al \shortcite{Langer:97a}, 
there may still be some effect on the surface abundances of Na,
O and Al as well as of the other lighter elements including such isotopes 
of Mg as $^{25}$Mg and $^{26}$Mg.

\begin{table}
\center
\begin{tabular}{|c|c|c|c|}
\hline

              &            	   &\textbf{Energy Generation}   &                   \\ 
\textbf{Case}&\textbf{Description}&\textbf{Reduction Factor $g$}&\textbf{Number of} \\
\textbf{}   &                    &\textbf{(of nuclear burning} &\textbf{Pulses}    \\
              &                    & \textbf{burning energy)}    &                   \\\hline
   0          & Normal Evolution   & 0.0                         & 0                 \\
   1          & Constant $g$       & 0.5                         & 1                 \\
   2          & Constant $g$       & 0.8                         & 1                 \\
   3          & Constant $g$       & 1.0                         & 1                 \\
   4          & Periodic pulse in $g$ 	   & 0.8                         & 20                \\
   5   	      & Periodic pulse in $g$	   & 0.5                         & 20                \\
   6          & Periodic pulse in $g$	   & 0.8                         & 20                \\
   7          & Periodic pulse in $g$ 	   & 1.0                         & 20                \\
   8          & Periodic pulse in $g$  	   & 1.5                         & 20                \\
\hline
\end{tabular}
\caption{\label{e_removal_models:tab} Reduced energy generation rate models.}
\end{table}

\begin{figure}
\includegraphics[width=0.9\textwidth, angle=0]{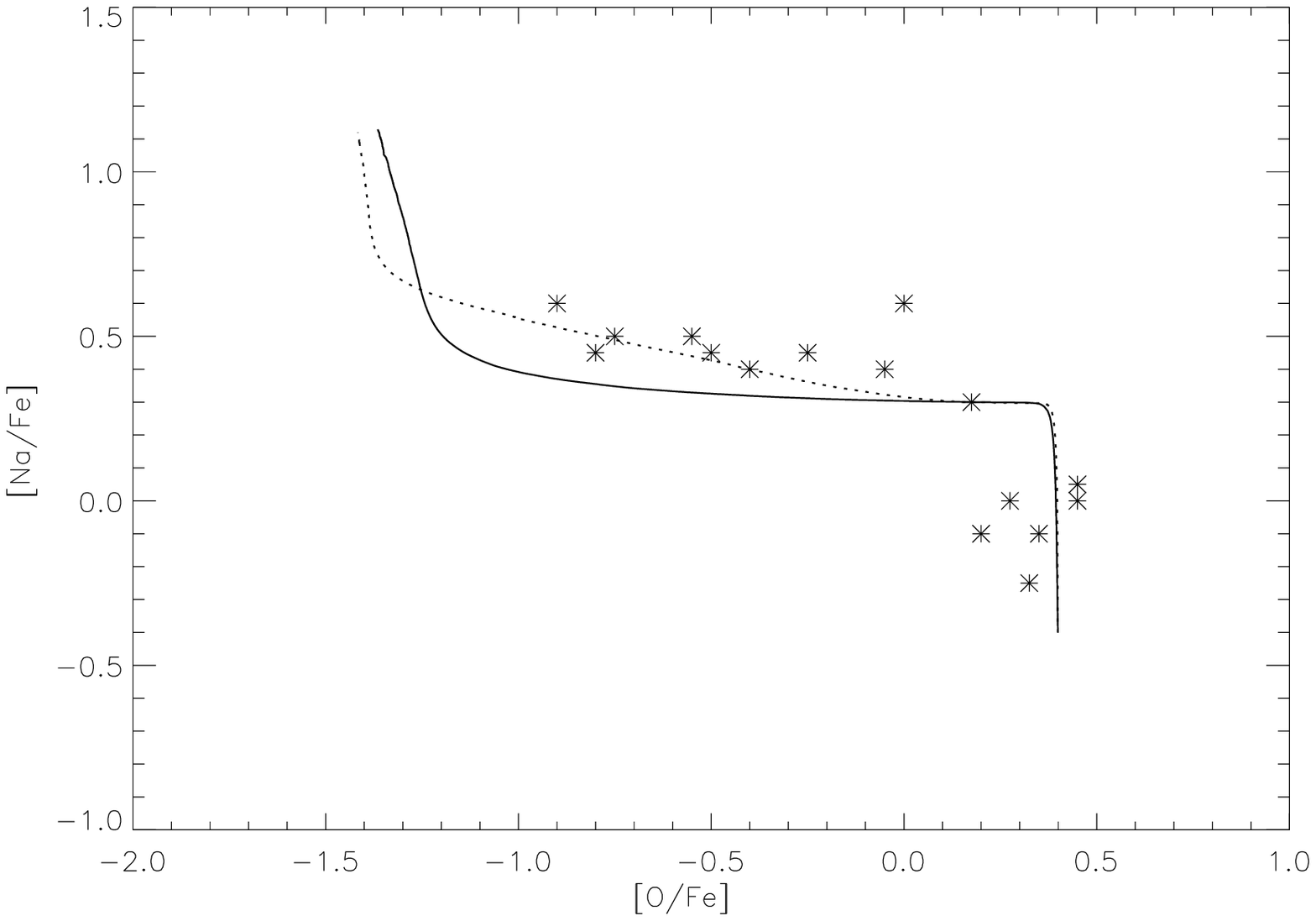}

\caption{\label{Na_time_pulse:fig} [Na/Fe] versus [O/Fe] for Case 6 in Table 1
(solid curve)
and a standard model with no energy generation rate modifications (dotted
curve, Case 0). The data points are for M13.}

\end{figure}

\begin{figure}
\includegraphics[width=0.9\textwidth, angle=0]{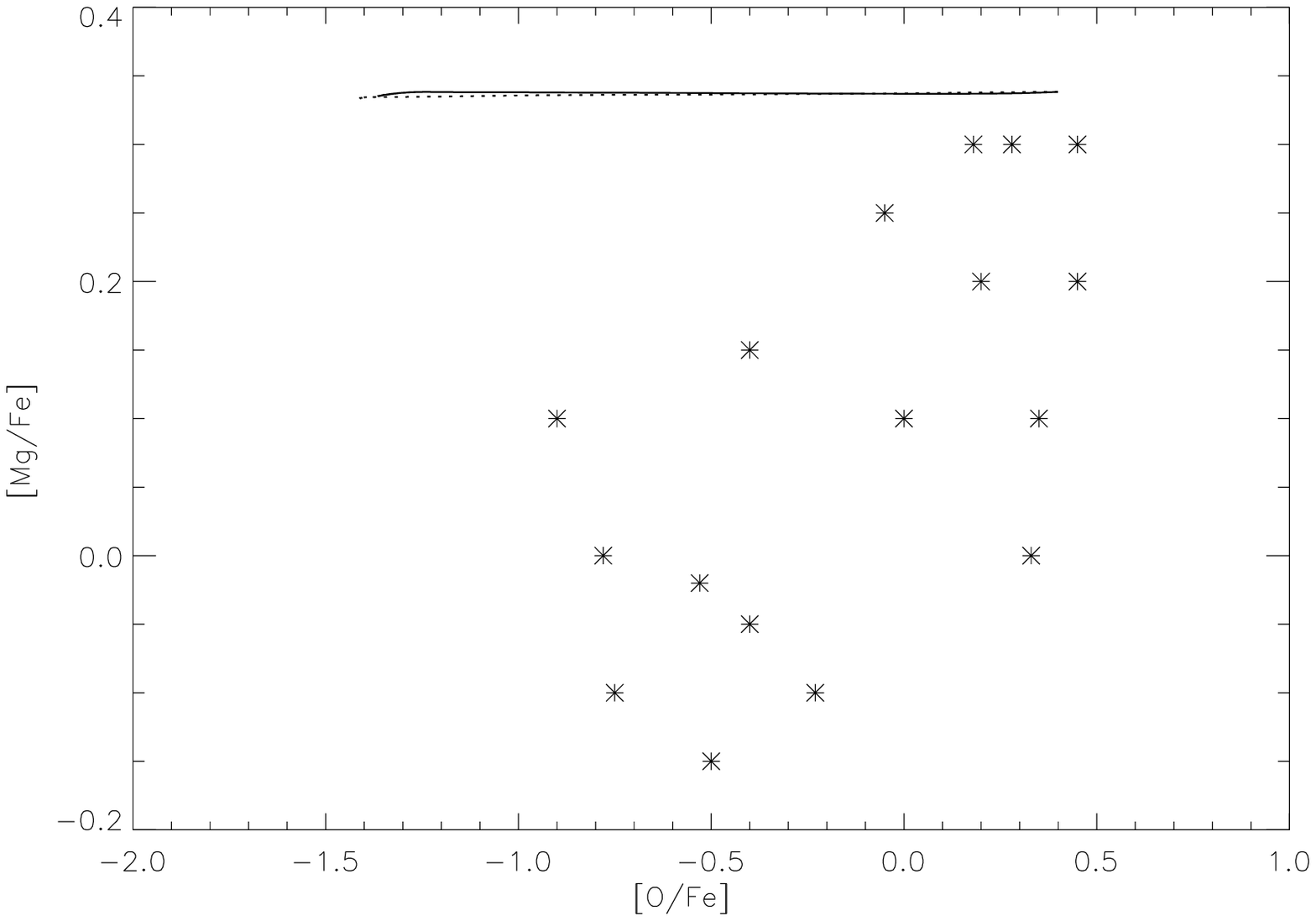}
\caption{\label{Mg_time_pulse:fig} Same as figure \ref{Na_time_pulse:fig} but for [Mg/Fe] versus [O/Fe].}
\end{figure}

\begin{figure}
\includegraphics[width=0.9\textwidth, angle=0]{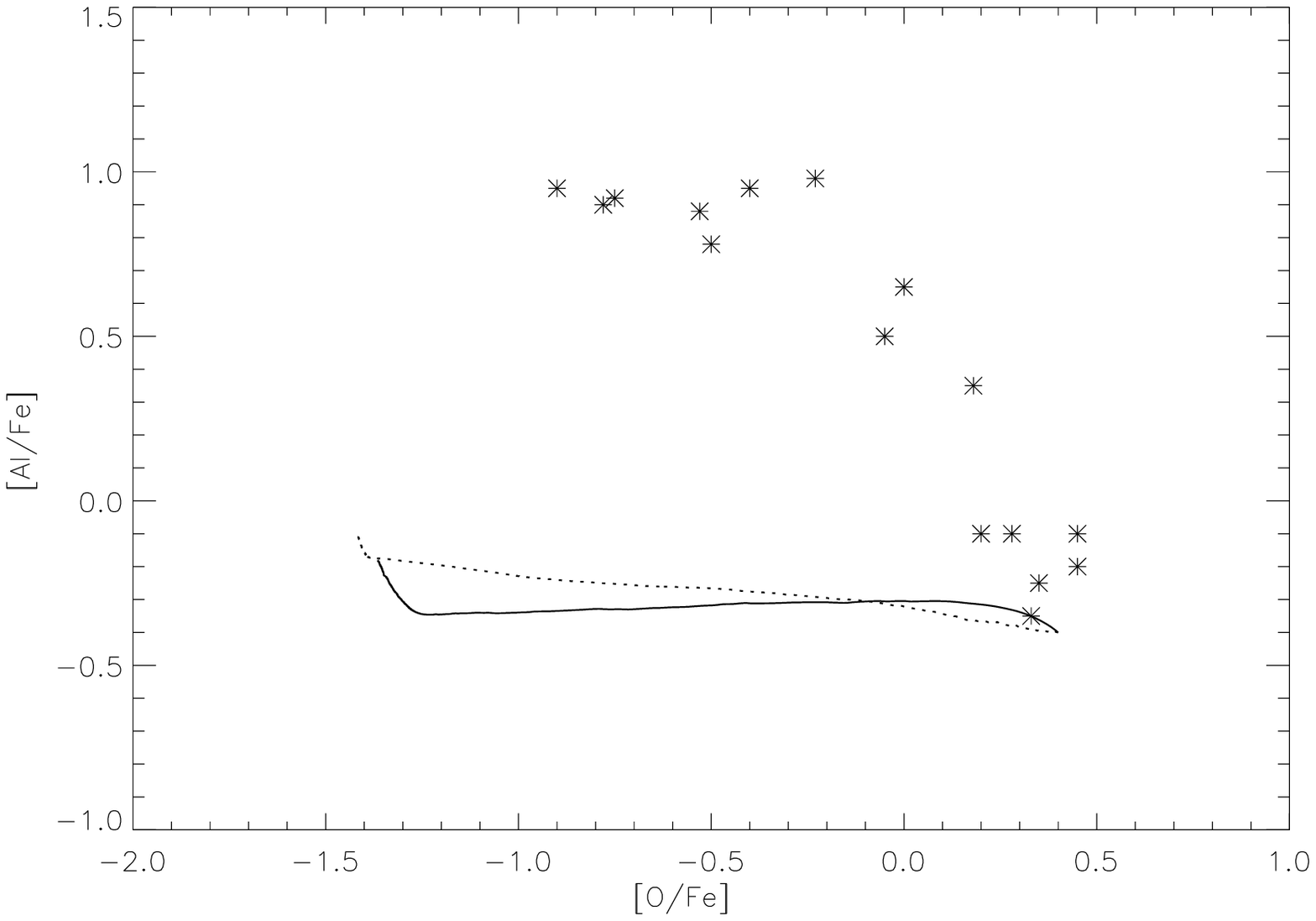}
\caption{\label{Al_time_pulse:fig} Same as figure \ref{Na_time_pulse:fig} but for [Al/Fe] versus [O/Fe].}
\end{figure}

Table \ref{e_removal_models:tab} lists all the models run where the 
energy generation rate was reduced. Case 6
was run through the nucleosynthesis code with deep mixing included.
Figures \ref{Na_time_pulse:fig},
\ref{Mg_time_pulse:fig} and \ref{Al_time_pulse:fig} show the surface abundances
for Na, Mg and Al
respectively with the curves for the standard model (Case 0) 
superimposed for comparison. It is quite
clear that the temperature pulses have had a negligible 
effect on the surface abundances.

\begin{figure}
\includegraphics[width=0.9\textwidth, angle=0]{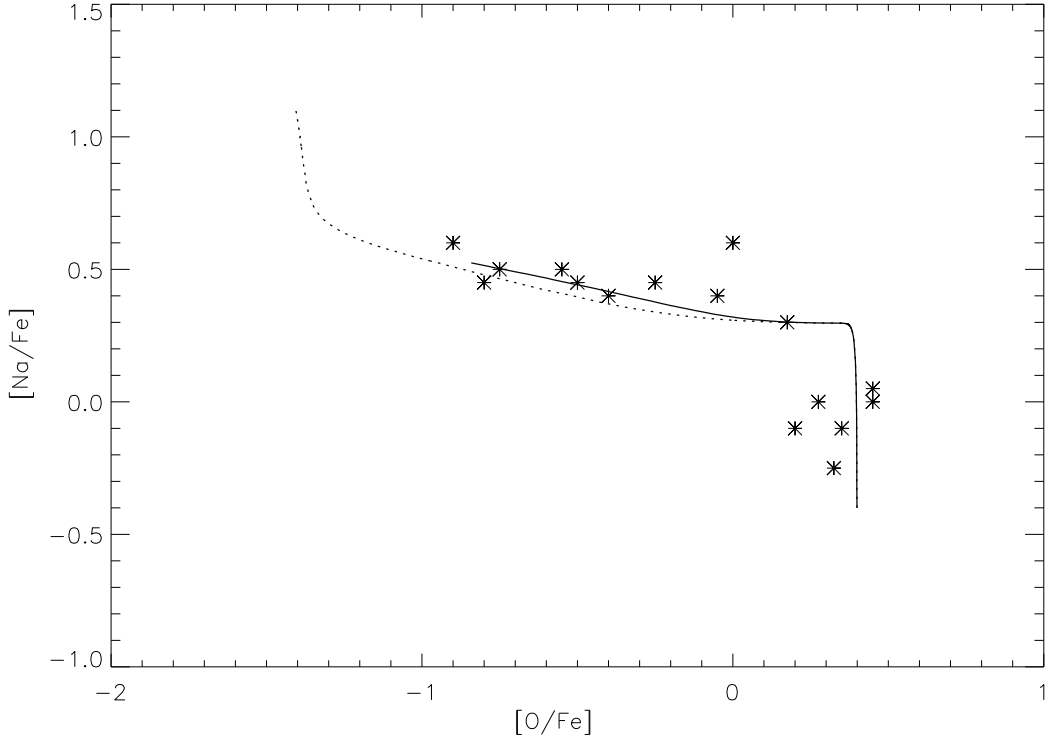}
\caption{\label{Na_const_removal:fig} [Na/Fe] versus [O/Fe] for Case 2
(solid curve) and a standard model with no
energy rate modifications (Case 0, dotted curve).}
\end{figure}

\begin{figure}
\includegraphics[width=0.9\textwidth, angle=0]{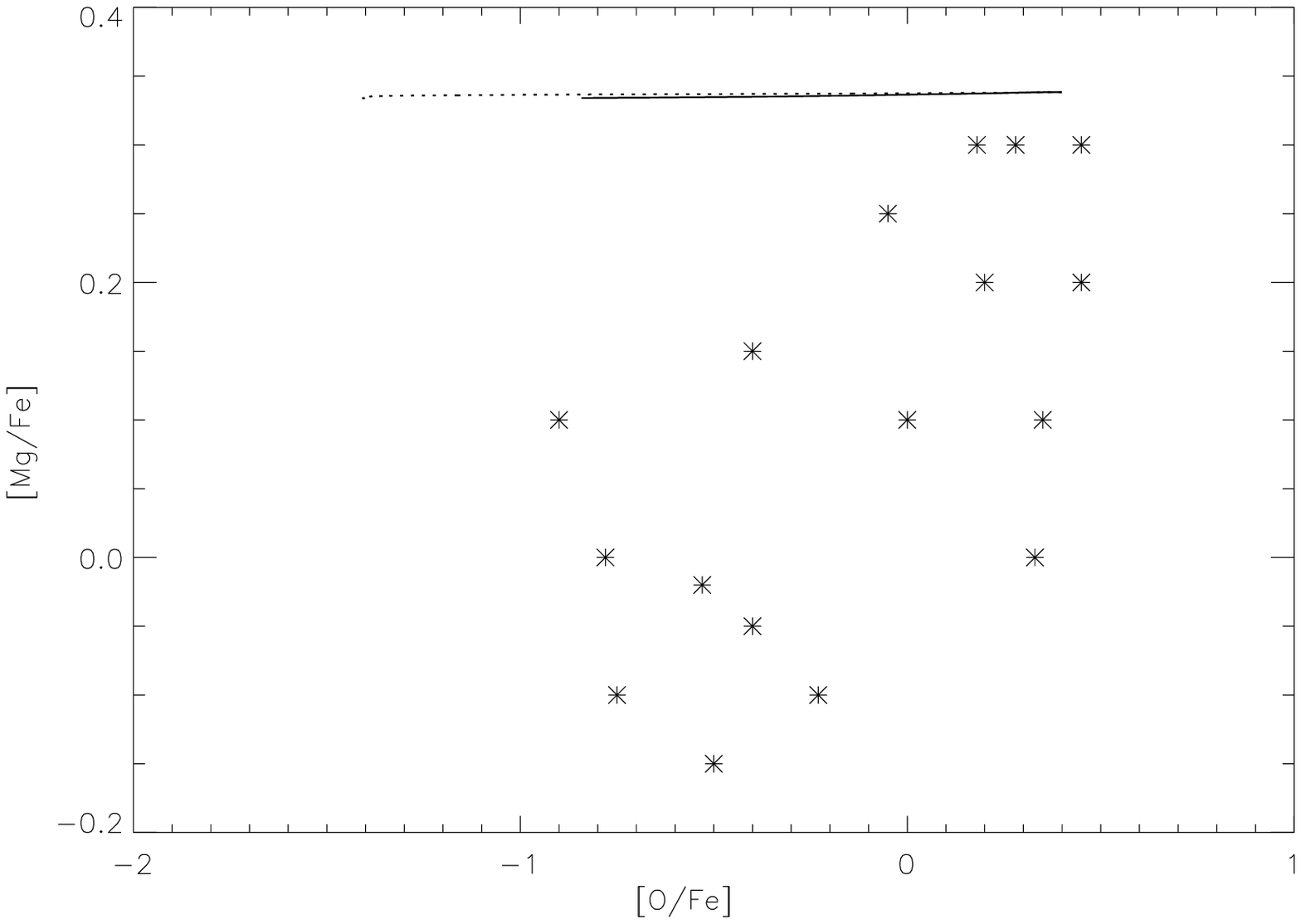}
\caption{\label{Mg_const_removal:fig} Same as figure \ref{Na_const_removal:fig} but for [Mg/Fe] versus [O/Fe].}
\end{figure}

\begin{figure}
\includegraphics[width=0.9\textwidth, angle=0]{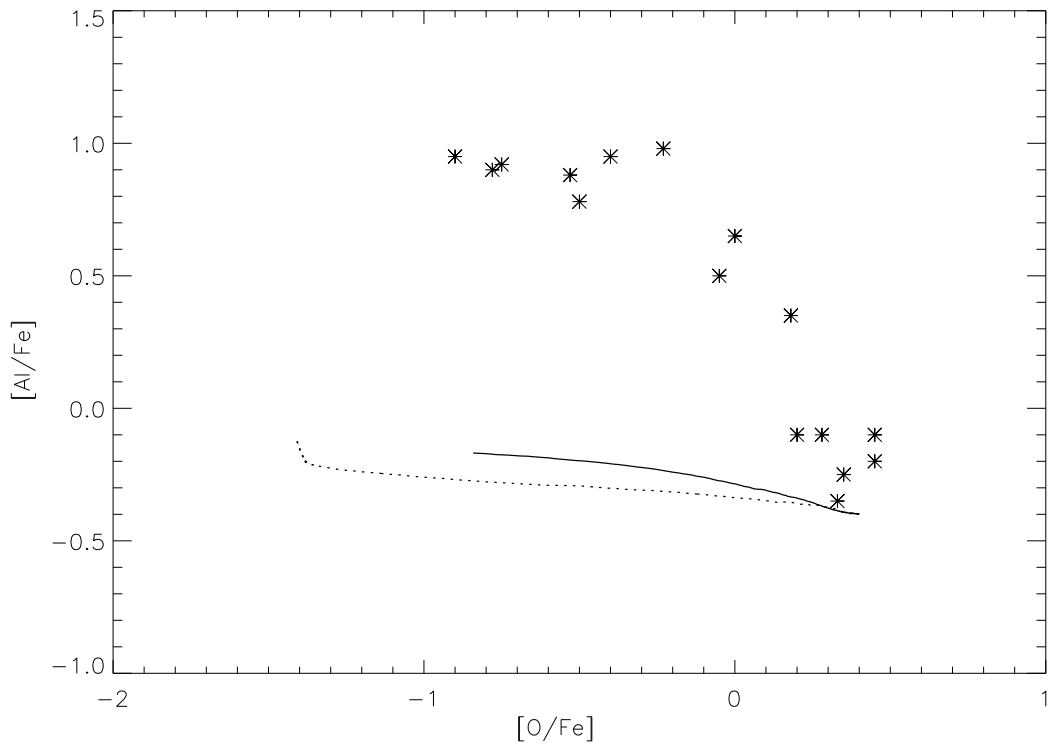}
\caption{\label{Al_const_removal:fig} Same as figure \ref{Na_const_removal:fig} but for [Al/Fe] versus [O/Fe].}
\end{figure}

The temperature rises are either too small or too short lived to have any
effect. Longer lived or more pulses only accelerate the
evolution of the star and does not assist in maintaining higher temperatures
and thus providing more $^{27}$Al for example. To test this, one of the models
with constant energy generation rate reduction, Case 2 from table
\ref{e_removal_models:tab} was also run through the nucleosynthesis code, with
its results and that of a standard model shown in figures
\ref{Na_const_removal:fig}, \ref{Mg_const_removal:fig} and
\ref{Al_const_removal:fig}. Once again there is little effect on the surface
abundances, as expected, although the Na values fit a little better.
The process accelerates the evolution but 
does not significantly alter the
surface abundances.

\section{Conclusion}
It appears that a thermally unstable H shell, modelled with the methods of this
paper,  cannot produce the high temperatures of about 70 million K
required to produce the observed aluminium abundances. 
We succeeded only in accelerating the evolution with negligible effect on the
interior temperatures and the surface abundances. Although a genuine instability,
may manifest itself in a way which is quite different to 
these exploratory calculations, we have tried to artificially 
reproduce the most favourable conditions without success.

It appears that primordial abundances variations among the stars combined with
deep mixing
currently offer greater potential in making progress on this problem.

\label{lastpage}

\end{document}